%% file: main.tex
\documentclass[aps,prd,reprint,secnumarabic,superscriptaddress,preprintnumbers,amsmath,amssymb,longbibliography,nofootinbib]{revtex4-2}

\usepackage{latexsym}
\usepackage{textcomp}
\usepackage{gensymb}
\usepackage{graphicx}
\usepackage{verbatim}
\usepackage{amsthm}
\usepackage{float}
\usepackage{amsmath}
\usepackage{booktabs}
\usepackage{graphicx,subfigure}
\usepackage{epstopdf}
\usepackage{indentfirst}
\usepackage{epsfig}
\usepackage{epsfig,subfigure,psfrag}
\usepackage{dcolumn}
\usepackage[T1]{fontenc}
\usepackage{bm}
\usepackage{slashed}
\usepackage{cancel}
\usepackage{xcolor}
\usepackage{tabularx}
\usepackage{hyperref}
\usepackage[normalem]{ulem}
\usepackage{braket}
\hypersetup{
    colorlinks=true,       
    linkcolor=blue,          
    citecolor=blue,        
    filecolor=blue,      
    urlcolor=blue           
}

\def\HEMT{\small{HEMT}}
\def\SPLENDOR{\small{SPLENDOR}}

\newcommand{\beq}{\begin{equation}}
\newcommand{\eeq}{\end{equation}}
\newcommand{\beqnarray}{\begin{eqnarray}}
\newcommand{\eeqnarray}{\end{eqnarray}}

\begin{document}
\title{Correcting impedance measurements for background parasitics to characterize circuit components in cryogenic environments}

\input{authors.tex}

\date{\today} 
\maketitle

{{\bf Abstract} - Predictable circuit response is a critical prerequisite for accurate electronic measurements. We describe a powerful, yet straightforward, experimental method and analysis model that utilizes an affordable LCR meter in conjunction with an \textit{in situ} parasitic impedance background correction procedure to measure the temperature-dependent impedance (magnitude and phase) of individual passive circuit elements mounted in a cryostat. 
We show how the model unambiguously identified a $\sim$~20x drop in capacitance for 22~$\mu$F 5XR multilayer ceramic capacitors cooled from 300~K to 360~mK in an environment with parasitic capacitance of $\sim$300 pF. The same experimental procedure, based on a simple two-wire measurement, was also used to successfully measure 10~pF and 22~pF thin-film capacitors and 100~M$\Omega$ thick-film resistors. The results showed that the resistor values increased by up to an order of magnitude when the devices were cooled from 300~K to 360~mK. Most importantly, the simple data acquisition method and robust analysis model were shown to effectively extend the accuracy of a commercial LCR meter beyond its manufacturer-guaranteed values for a wide range of measurement frequencies.}

\section{Introduction}
\label{Intro}
The successful implementation of cryogenic dark matter detectors necessitates coupling the detectors to low-noise signal amplification circuits that enable detector readout with well-controlled noise mitigation. As has been well established in the field of low-temperature detectors, the cryogenic environment itself complicates typical amplification methods because the electrical properties of typical materials depend on the temperature~\cite{baulieu}. Therefore, in order to obtain detector data with an excellent signal-to-noise ratio, circuit components must be carefully selected to ensure their reliable operation at cryogenic temperatures. 

\begin{figure}
\center
\includegraphics[width=3.2 in, keepaspectratio]{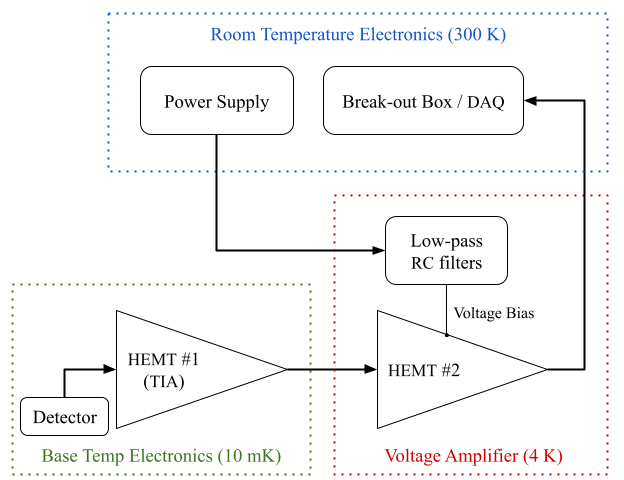}   
\caption{Simplified circuit schematic of the two-stage cryogenic \HEMT-based amplifier designed for SPLENDOR~\cite{anczarski}. A base-temperature board serves as a ``buffer'' stage that connects the first stage of the amplifier to the cryogenic detector through a coupling capacitor. This first stage essentially serves as a transimpedance amplifier (TIA). A second-stage board at 4~K implements a second {\HEMT} to amplify the detector signal for room temperature readout. Low-pass filters on the 4~K stage are used to help mitigate high frequency noise on the {\HEMT} bias lines. (Additional  filters at 300~K not shown.) Measuring the explicit temperature dependence of the specific capacitors and resistors used for this amplifier system was a primary goal of the work reported here.}
\label{fig:newBlockDiagram}
\end{figure}

For components such as silicon-based transistors, low operating temperatures can result in ‘freeze-out’ behavior, where charge carriers no longer exist in sufficient quantities within the conduction band~\cite{juillard}. 
High electron mobility transistors ({\HEMT}s) solve the freeze-out problem and offer an alternative means of signal amplification at 4~Kelvin~\cite{phipps}. The {\SPLENDOR} collaboration has recently developed a novel two-stage cryogenic {\HEMT} amplifier scheme \cite{anczarski} that first converts a charge signal to a voltage signal in a 10~mK buffer stage and then amplifies the voltage signal in a separate 4~K gain stage (see Fig.~\ref{fig:newBlockDiagram}). This amplified signal is brought to 300~K for further amplification and filtering. 

Although the {\HEMT}-based amplifier scheme can provide excellent energy sensitivity, the system is still susceptible to extraneous electrical noise, including noise on the cryogenic {\HEMT} bias and signal readout lines~\cite{anczarski}. While using simple RC filters can help mitigate such noise, most resistors and capacitors exhibit significant temperature dependence, so their use in precision cryogenic measurement schemes can become problematic. As a result, we developed a systematic experimental method to accurately and precisely measure the complex impedance of circuit components such as filter resistors and capacitors as a function of temperature, using a simple and affordable bench-top LCR meter. Below, we provide an overview of our experimental setup and describe the analytical model we developed to extract key component values from raw data that contain substantial parasitic contributions. We show how this new model is ultimately limited by the performance of our commercial LCR meter. We conclude with a summary of our results so far and our plans for using this new experimental tool in the near future for important measurements.

\section{Methods}
\label{Methods}
For the experiments described in this paper, the magnitude and phase of the impedance ($\mathbb{Z}_x$) of eight test components were seprately measured with a bench-top B\&K Precision 891 LCR meter over a frequency range of 20 Hz to 300 kHz, using logarithmic spacing. Two separate frequency scans were performed for each component, at three different temperatures (room temperature, 12 K, and 360 mK), to independently collect phase and magnitude information. All frequency sweeps used a fixed root-mean-square voltage signal of 1.0 V.

\begin{figure}
\centering
\includegraphics[width=2.4in, keepaspectratio]{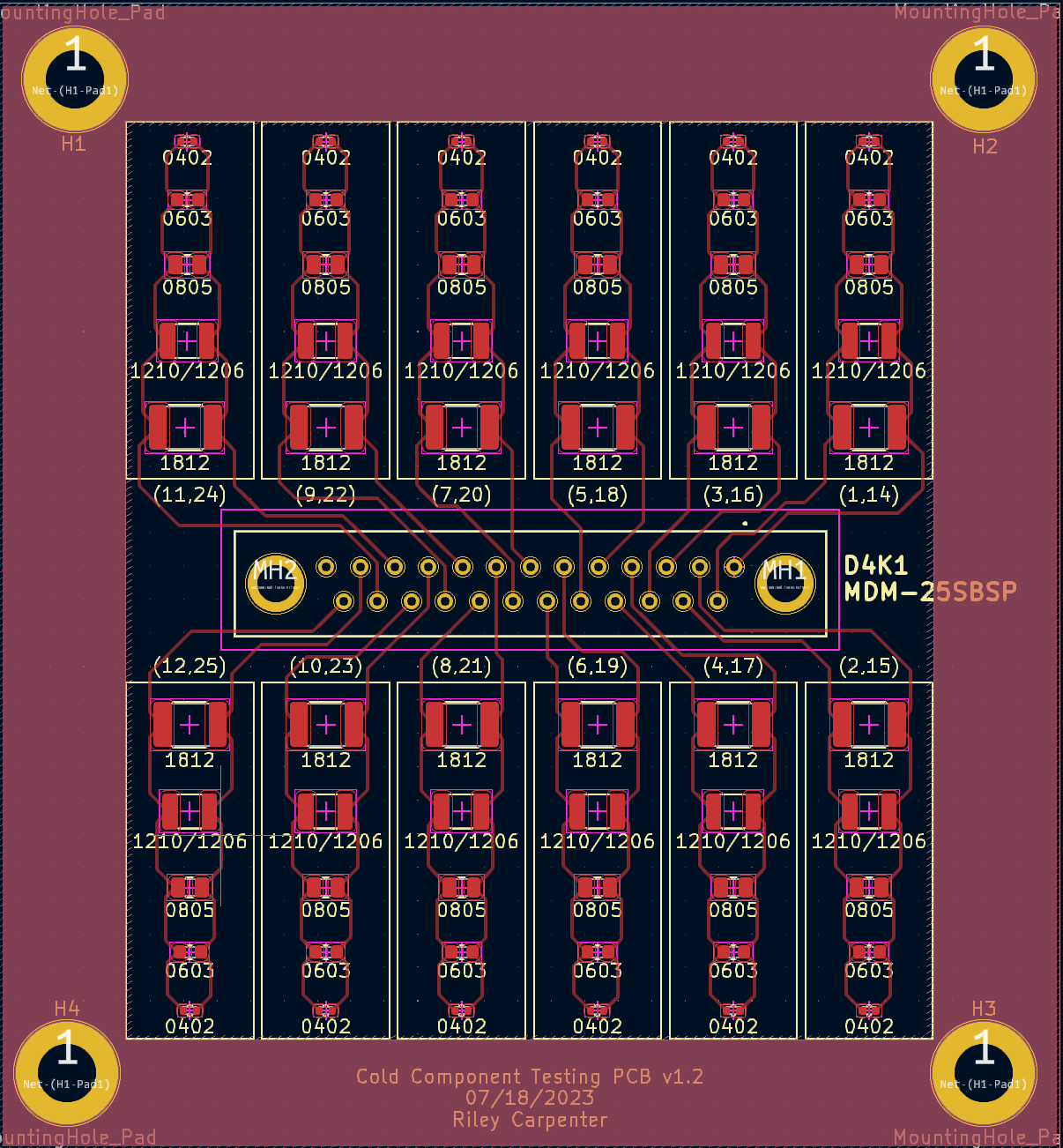}
\caption{Multi-layer printed circuit board (PCB) layout v1.2 of the 12-channel cold component test mount used for this work. The board is 80 mm wide x 86 mm tall. Channel numbers run from right to left by column: Ch 1 is in the upper right corner, Ch 2 in the lower right corner, and so on, ending with Ch 12 in the lower left corner. In our experiment, Ch 7 and Ch 11 served as designated ``open" channels and Ch 8 and Ch 12 were  ``shorted" using 0 $\Omega$ resistors with a 1206 footprint.}
\label{fig:PCB_Layout}
\end{figure}

The components under test were soldered directly to copper pads on the printed circuit board (PCB) shown in Fig.~\ref{fig:PCB_Layout}. The board was designed to connect twelve single-component circuit paths to twelve independent twisted-pair channels via an MDM-25 connector. In practice, 8-10 components are tested in a given cryogenic run; the remaining channels serve as \textit{in situ} ``open'' and ``short'' calibration channels. These calibration channels allow for the parasitic impedance of the cryostat wiring, PCB, and room temperature break-out box to be systematically subtracted from the raw impedance measurements of test components. 

Each PCB channel can accommodate a surface mount component with one of six common footprints (1812, 1210, 1208, 0805, 0603, 0402) for flexible testing compatibility. An open calibration channel is identical in layout to the others and simply does not have a component soldered onto any of the pads. Similarly, a shorted channel is defined by soldering a 0 $\Omega$ jumper (1206 footprint) in place of a test component. Since the relative position of each channel can impact parasitic impedance values, the experimental platform was designed with several options for choosing specific open/short calibration channels, such that each remaining test channel could be calibrated with an open/short pair in a similar location on the PCB.

 The LCR meter itself was also calibrated $in~situ$ by performing  two sets of impedance scans: first, the LCR probe cables were disconnected from any input source (or each other) at the far end (open-circuit measurement). Then, the probe cables were connected directly to each other (short-circuit measurement). This meter calibration was executed only once per cryostat cool-down.

Although the LCR meter supports four-wire measurements, a two-wire measurement scheme was used inside the cryostat to maximize the number of test samples that could be studied in each cool-down. Thus, natural changes in impedance of the circuit wiring, including twisted-pair constantan and twisted-pair superconducting NbTi wire looms, 
directly impacted LCR meter readings. These expected sources of parasitic impedance were addressed using the on-board calibration capabilities (``open'', ``sort'' measurements) and $insitu$ calibration impedance scans  described above. For the experiments described here, two component positions (Chs 7, 11) on the sample PCB were designated as ``open'' channels (no component installed) and two others (Chs 8, 12) were selected as ``short'' channels, where a 0~$\Omega$ resistor was mounted in place of a test component. Using data from these two channels in combination with the $in\, situ$ meter calibration data, enabled accurate accounting for variations in channel-specific parasitic behavior for every physics measurement. 

The remaining eight channels of the test board were populated with the devices under test, including four 100~M$\Omega$ thick film resistors, two 22~$\mu$F multilayer ceramic capacitors, and two thin-film capacitors with values of 22~pF and 10~pF.  Table~\ref{tab:componentTbl} summarizes the specific channel assignments for all 12 PCB channels used in the experimental run. The 22~$\mu$F capacitors were identical to 
the 22~$\mu$F capacitors used for the {\SPLENDOR} experiment. 
Each measurement scan of $|\mathbb{Z}|$ (or $\theta$) versus frequency was saved directly from the LCR meter as a text file. The raw data files were then parsed for non-numerical entries, converted to base SI units, and imported into {\small MATLAB} for subsequent processing and analysis.


\begin{table*}
\small
\centering
\caption{\label{tab:componentTbl} List of test component properties and test board channel numbers for the first experimental run. The manufacturers' minimum temperature specifications for these components is -55$^o$C.}
\

\begin{tabular}{| c | c | c | c | c | c | c |}
\hline
Ch. & Nom. &  Tol. & Temp. Var. & Footprint & Type & Manufacturer \\  
 & Value & & (ppm/$^o$C) & & & \\
\hline
1 & 100 M$\Omega$ & $\pm$1\% & $\pm$50 & 1206 & Thick-film & Stackpole Electronics \\
\hline
2 & 22 pF & $\pm$1\% & $\pm$30 & 0805 & Thin-film & Kyocera AVX \\
\hline
3 & 100 M$\Omega$ & $\pm$1\% & $\pm$100 & 0603 & Thick-film & Ohmite \\
\hline
4 & 100 M$\Omega$ & $\pm$5\% & $\pm$300 & 0805 & Thick-film & Yageo \\
\hline
5 & 10 pF & $\pm$1\% & $\pm$30 & 0402 & Thin-film & Kyocera AVX \\
\hline
6 & 100 M$\Omega$ & $\pm$1\% & $\pm$50 & 1206 & Thick-film & Stackpole Electronics \\
\hline
7 & $\infty$ $\Omega$ & -- & -- & -- & -- & -- \\
\hline
8 & 0 $\Omega$ &  50 m$\Omega$ & -- & 1206 & Thick-film & Panasonic\\
\hline
9 & 22 $\mu$F & $\pm$10\% & 5XR & 1812 & Multilayer ceramic & Kyocera AVX \\
\hline
10 & 22 $\mu$F & $\pm$10\% & 5XR & 1812 & Multilayer ceramic & Kyocera AVX \\
\hline
11 & $\infty$ $\Omega$ & -- & -- & -- & -- & -- \\
\hline
12 & 0 $\Omega$ & 50 m$\Omega$ & -- & 1206 & Thick-film & Panasonic\\
\hline
\end{tabular}
\end{table*}
\

\section{Experimental Model}
\label{Experimental Model}

\begin{figure}
\centering
\includegraphics[width=3.4 in]{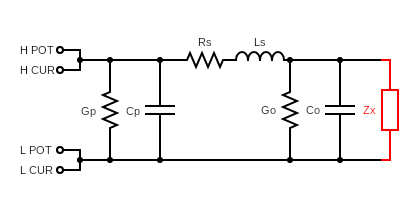}
\caption{Circuit model used to determine parasitic contributions to LCR meter impedance measurements of a test device $Z_x$ (in red). Twelve devices can be tested in a single cryostat run. The PCB open-circuit calibration measurement is represented by having $Z_x\rightarrow\infty$; the short-circuit measurement uses $Z_x\rightarrow 0~\Omega$. (See text.)}
\label{fig:CircuitDiagram}
\end{figure}

In this section, we describe a robust analytical model and analysis method developed to reliably extract sample resistance and capacitance values from complex cryogenic data sets containing substantial and variable parasitic contributions. 
Using this model, depicted schematically in Fig.~\ref{fig:CircuitDiagram}, the measured impedance, $\mathbb{Z}_m$, of a passive component such as a resistor or capacitor can be expressed as a combination of the true, temperature-dependent component impedance, $\mathbb{Z}_x$, parasitic series impedances ($R_s$, $L_s$), and parallel admittances ($G_p$, $C_p$, $G_o$, $C_o$).  The parasitic contributions are determined explicitly using data collected for dedicated PCB-level ``open'' and ``short'' calibration channels (see Section II), thereby enabling $\mathbb{Z}_x$, the true, isolated component impedance to be expressed in terms of measured quantities. 

The parasitic impedance in series with the test component models the parasitic resistance and inductance in the cryostat wiring and PCB traces. This combined quantity is denoted by:
\begin{equation}
{\mathbb{Z}_{s} = R_{s}+j\omega L_{s}},
\label{eq:ZsDef}
\end{equation}
where $\omega$ = 2$\pi$ x sampling frequency.
Similarly, the circuit model shown in Fig. 3 includes a parasitic capacitance and conductance in parallel with the test component on the PCB. This combined ``admittance" term contributes to the open channel calibration but not the shorted one, and is denoted by:
\begin{equation}
{\mathbb{Y}_o = G_o + j\omega C_o}.
\label{eq:YoDef}
\end{equation}
Lastly, there is an additional parasitic admittance originating from two coaxial BNC cables connecting the 4-wire LCR meter terminals to a 2-wire breakout-box connection to the cryostat. This admittance is in parallel with the rest of the circuit and is denoted in our model by:
\begin{equation}
{\mathbb{Y}_p = G_p + j\omega C_p}.
\label{eq:YpDef}
\end{equation}

The three measured impedance quantities, $\mathbb{Z}_{Sh}$, $\mathbb{Z}_{Op}$, and $\mathbb{Z}_{m}$, corresponded to the short, open, and test component configurations respectively. For $\mathbb{Z}_x$ to be expressed fully in terms of these three quantities without introducing free parameters, this model was approximated by a nominal pi-model for a transmission line by assuming $\mathbb{Y}_p=\mathbb{Y}_o$. Under this approximation, the three measured quantities obey the following relations:
\begin{equation}
{\frac{1}{\mathbb{Z}_{Sh}} = \mathbb{Y}_o + \frac{1}{\mathbb{Z}_s}}
\label{eq:ZshDef}
\end{equation}
\begin{equation}
{\frac{1}{\mathbb{Z}_{Op}} = \mathbb{Y}_o + \frac{1}{\mathbb{Z}_s + \frac{1}{\mathbb{Y}_o}}}
\label{eq:ZopDef}
\end{equation}

\begin{equation}
{\frac{1}{\mathbb{Z}_{m}} = \mathbb{Y}_o + \frac{1}{\mathbb{Z}_s + \frac{1}{ \mathbb{Y}_o + \frac{1}{\mathbb{Z}_x}}}}
\label{eq:ZmDef}
\end{equation}

By algebraically manipulating Eqs.~\ref{eq:ZshDef}-\ref{eq:ZmDef}, the parasitics-corrected, true test component impedance $\mathbb{Z}_{x}$ can be expressed in terms of the measured impedance values only -- corresponding to PCB channel ``open'' ($\mathbb{Z}_{Op}$), PCB channel ``shorted'' ($\mathbb{Z}_{Sh}$) and PCB channel with test component connected ($\mathbb{Z}_{m}$):

\begin{equation}
\mathbb{Z}_x = \frac{\mathbb{Z}_m - \mathbb{Z}_{Sh}}{1 - \frac{\mathbb{Z}_m}{\mathbb{Z}_{Op}} }.
\label{eq:modelEqn1}
\end{equation}

\begin{figure}
\centering
\includegraphics[width=3.4 in]{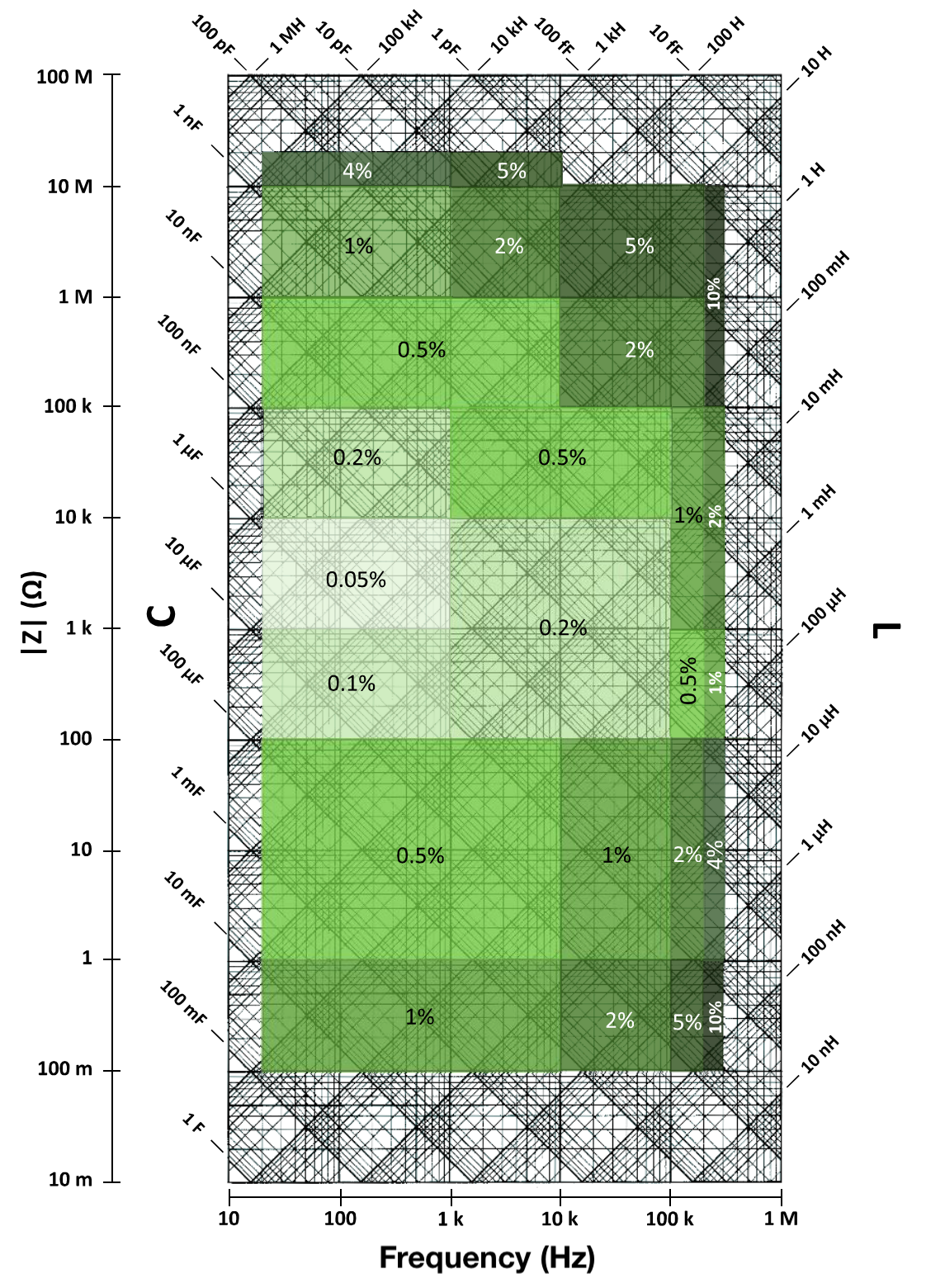}
\caption{Measurement accuracy chart shown in the user manual of the B\&K Precision 891 LCR meter~\cite{lcrManual}.}
\label{fig:LCRspecs}
\end{figure}

The LCR meter used enables measurements from 20~Hz to 300~kHz. However, its accuracy specifications depend on both $|\mathbb{Z}|$ and the measurement frequency, as shown in Fig.~\ref{fig:LCRspecs}. In particular, the meter is not rated for impedance magnitudes $>$~10 M$\Omega$, except in the 20~Hz-10~kHz region, where the impedance specification extends to 20~M$\Omega$.
It is critical to note, however, that the impedance accuracy limitations apply to \textit{measured} impedance values only, in our case: $\mathbb{Z}_{Sh}$, $\mathbb{Z}_{Op}$ and $\mathbb{Z}_m$. In particular, owing to the various parasitic effects that reduce impedance mentioned above, we showed that the true impedance of a test component, $\mathbb{Z}_x$, could exceed the upper bound in magnitude without any single measurement pushing the LCR meter beyond its quoted limits.
\section{Results and Discussion}
\label{Results and Discussion}
\subsection{Open and Short Parasitics Measurements} 
\label{Open/Short}
\begin{figure}
\centering
\includegraphics[width=3.4 in]{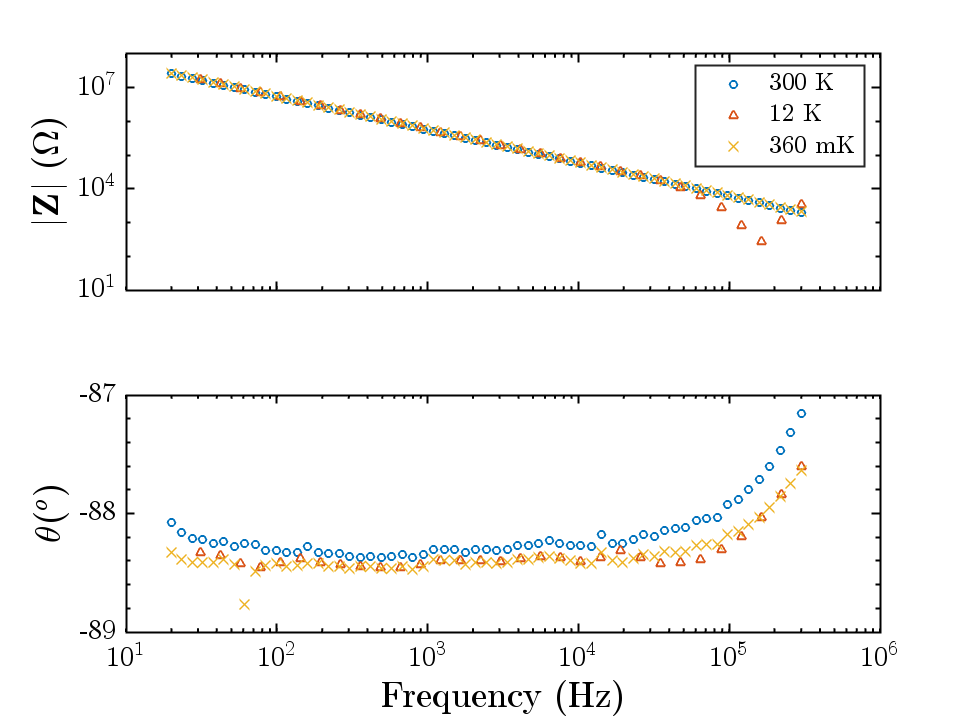}
\caption{Magnitude and phase data $\mathbb{Z}_{Op}$ for ``open'' calibration channel (Ch. 7). Visually, the impedance magnitudes were relatively independent of temperature, although the phase did shift towards -90° for the lower-temperature data. A single 60~Hz outlier point is shown for completeness and is attributed to sporadic noise pick-up in the lab.}
\label{fig:OpenData}
\end{figure}

Impedance measurements for the open channels at all three temperatures (300 K, 12 K and 360 mK) exhibited capacitive behavior, with a phase angle $\theta$ close to -90$^o$ that started to diverge towards smaller negative values at frequencies above $\sim$100 kHz (see Fig.~\ref{fig:OpenData}). The frequency dependence of the impedance magnitudes for the open channels was similar at all three temperatures. Although the measured phase angle was slightly larger for the 300 K data relative to the 12 K and 360 K data, the relative difference was small enough to be limited by the known accuracy of the LCR meter~\cite{lcrManual}.
However, it was surprising that the phase angle did not generally exceed approximately -88.5$^o$, despite the impedance circuit in concept being predominantly capacitive, especially in the intermediate frequency regime near 1 kHz. In fact, across all 12 channels at all three measurement temperatures, the two most negative phase values recorded directly by the LCR meter were -88.77$^o$ and -88.68$^o$, measured at a frequency of 61.4 Hz and T = 360 mK on the open calibration channels 7 and 11.

As shown for Ch. 7 in Fig.~\ref{fig:OpenData}, this point at $\sim$ 60 Hz appeared to be an outlier compared to a relatively stable average phase angle between 30 Hz and 10 kHz. Over this range, the mean phase angle was -88.43$\pm$0.07$^o$, making the point near 60 Hz more than 5~$\sigma$ away from the local mean. Similarly, for the open calibration data on Ch. 11, the average phase angle was -88.47$\pm$0.05$^o$ for the 360 mK measurements, making this outlier of -88.68$^o$ $\sim$ 4~$\sigma$ from the mean. Room temperature data on Ch. 11 also exhibited a phase outlier more than 3.5~$\sigma$ below the mean, again at 61.4 Hz. The logarithmic sampling distribution in frequency for these scans had a spacing of roughly 10 Hz between points in this range, making these data points the closest in frequency to 60 Hz, a common noise frequency in most laboratories. 
These phase angle outliers in the data appear correlated with the proximity of the test frequency to 60 Hz, which in turn lends credence to a universal calibration error in the LCR meter phase data. During the open calibration of the LCR meter itself, prior to PCB data collection, the BNC cable leads from the meter were held apart, but did not have their ends capped. Therefore, stray 60 Hz noise from outlets and power strips in the lab may have caused a similar negative spike in the phase angle measured while sweeping past 60 Hz. Such a negative outlier would have been used by the meter's internal calibration protocol to set the outermost phase angle limit of -90$^o$, thereby resulting in subsequent measurements of theoretically capacitive systems to be registered as less extreme, since the 60 Hz outlier point set an otherwise unmet -90$^o$ limit.

Although not ideal, the observation and \textit{post hoc} identification of this unexpected phase offset is not without value. For one, it affirms the potential for extraneous noise during the LCR meter calibration to cause unexpected shifts in the LCR meter measurements. Since the parasitic environment for the PCB is known to change between temperatures, relying only on the LCR meter calibration and remeasuring the open/short background for each temperature could introduce systematic variations. Conversely, sticking to one universal LCR meter calibration with additional on-board calibration infrastructure ensures that unpredictable noise sources such as stray 60 Hz signals in the ambient lab environment will appear as a consistent systematic offset, apparent across all channels. For the purposes of this analysis, a correction model for this observed offset was not implemented, since the point-by-point impedance transformation defined in Eq.~\ref{eq:modelEqn1} did not require $\mathbb{Z}_{Op}$ to be purely capacitive, suggesting that this model could account for such a universal calibration offset when used to calculate component impedance values.

\begin{figure}
\centering
\includegraphics[width=3.4 in]{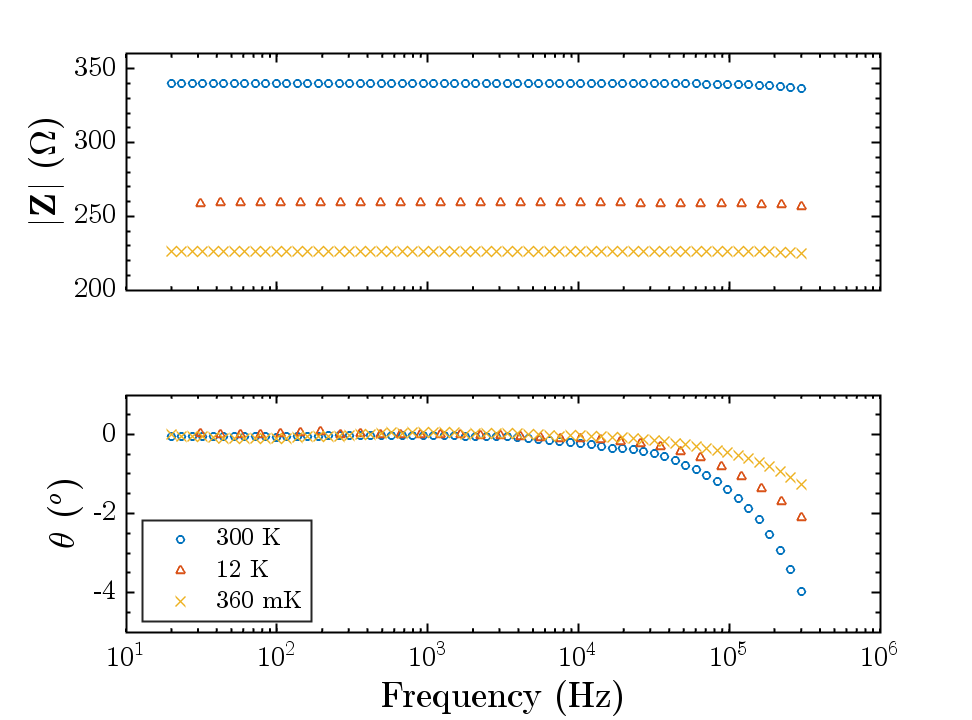}
\caption{Magnitude and phase data $\mathbb{Z}_{Sh}$ for a ``short'' calibration channel measurement (Ch. 8). The drops in impedance between temperatures are associated with the superconducting transition of the cryostat wiring connecting the LCR meter to the PCB.}
\label{fig:ShortData}
\end{figure}

At each temperature, the measured impedance magnitude of the ``short'' channels was generally independent of frequency, corresponding to predominantly resistive behavior (see Fig.~\ref{fig:ShortData}). The onset of a phase angle divergence in the negative direction at frequencies past $\approx$10-100 kHz suggests that the parasitic parallel capacitance $C_p$ has a stronger effect on the overall impedance than $L_s$, at least for frequencies below 300 kHz.
\subsection{Frequency Interpolation}
\label{Interpolation}
\begin{figure}
    \centering
    \includegraphics[width=3.4 in]{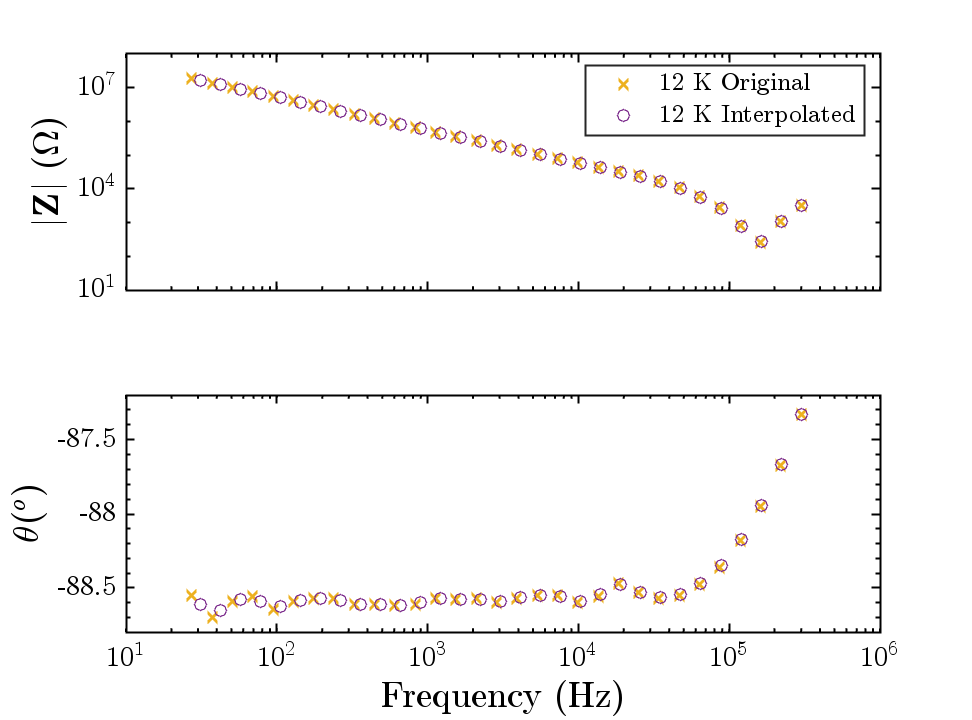}
    \caption{Example of the transformation between raw data (\textcolor{brown}{x}) and interpolated data (\textcolor{violet}{o}) for sample Ch.~2 (22 pF) measured at 12~K. The interpolated data points are aligned to the frequencies sampled for Ch.~11 (open circuit) at 12~K.}
    \label{fig:Ch2_interp}
\end{figure}

For certain frequency sweeps, the automatic scaling setting for the LCR meter was enabled, resulting in channel-to-channel differences in the sampled frequency values. For the impacted ($f$,$|\mathbb{Z}|$) and ($f$,$\theta$) data sets (identified in Table~\ref{tab:interpInfo}), the measured data were systematically interpolated to align the sampled frequencies across all twelve channels at each temperature ($e.g. $ see Fig.~\ref{fig:Ch2_interp}). This step was necessary to ensure that the impedance correction method used values of $\mathbb{Z}_m$, $\mathbb{Z}_{Sh}$, and $\mathbb{Z}_{Op}$ corresponding to self-consistent frequencies. 
\begin{table*}
\small
\caption{\label{tab:interpInfo}List of channels for which interpolation was used to systematically align frequencies between channels at each testing temperature. The open calibration line on Ch. 11 was selected for the standardized frequency points. Those entries marked with an 'x' required interpolation to convert to frequency-aligned data sets. Unmarked entries did not undergo any interpolation.}
\centering
\begin{tabular}{| c | c | c | c | c | c | c | c | c | c | c | c | c |}
\hline
Ch. & 1 & 2 & 3 & 4 & 5 & 6 & 7 & 8 & 9 & 10 & 11 & 12 \\  
\hline
300 K & & & & & & & & & & & n/a & \\
\hline
12 K & x & x & & & & & x & x & x & x & n/a & x \\
\hline
360 mK & x & x & x & & & & & & & & n/a & \\
\hline
\end{tabular}
\end{table*}

\subsection{Test Capacitor Measurements}
\label{Capacitors}
The parasitics-corrected, actual capacitance value $C_x$ for each test capacitor:
\begin{equation}
    {C_x = -\frac{1}{2\pi f *\mathfrak{Im}(\mathbb{Z}_x)}}
    \label{eq:CxDef}
\end{equation}
was calculated as a function of frequency, in a point-wise manner, using Eq.~\ref{eq:modelEqn1} to transform as-measured impedance values $\mathbb{Z}_m$ into  isolated component values $\mathbb{Z}_x$. Instances where the corrected capacitance was relatively stable in the low- to mid-frequency range were indicative of successful component isolation dyring the analysis. Based on graphical inspection, a suitable frequency range was determined for each test capacitor over which to average the capacitance and determine its standard deviation (see Tbl.~\ref{tab:CapacitanceTable}).

\begin{figure}
\includegraphics[width=3.4 in]{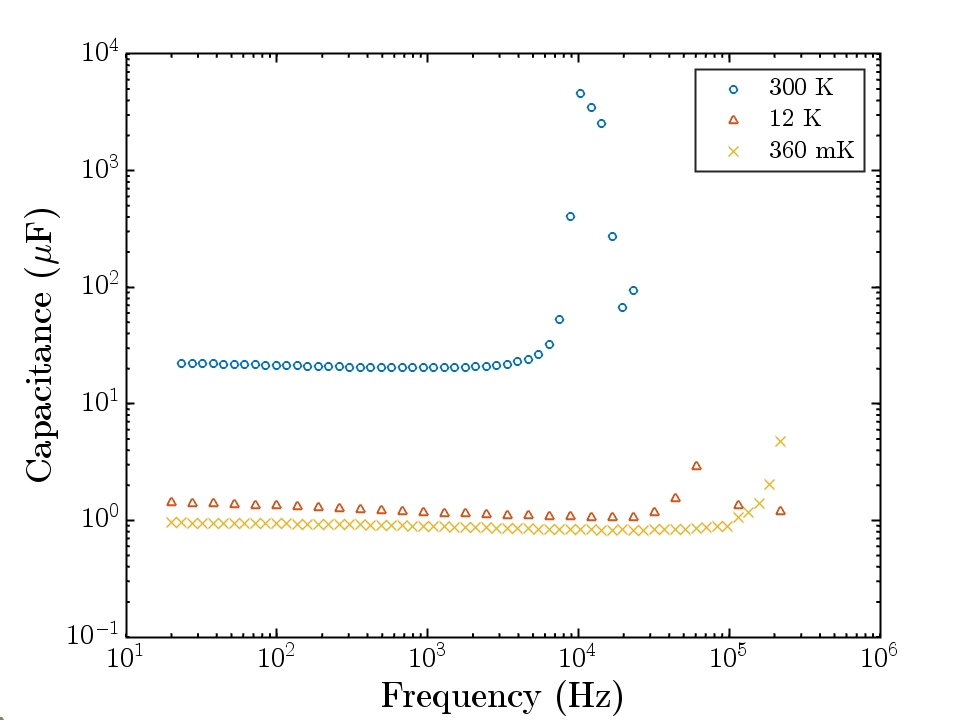}  
\caption{Parasitics-corrected component capacitance vs. frequency for the nominally 22~$\mu$F 5XR dielectric multilayer ceramic capacitor on Ch. 10, using the Ch.~7 ``open'' and Ch.~8 ``short'' calibration lines for component correction. At 300 K (\textcolor{blue}{o}), the low-frequency capacitance value was 21.6~$\pm$~0.9 $\mu$F (95\% CL), well within the 10\% tolerance specified by the manufacturer.  By 12~K, the capacitance had decreased to 1.37~$\pm~$0.09~$\mu$F. At the cryostat base temperature of 360 mK (\textcolor{brown}{x}), the value was only 940~$\pm$~18 nF, a $\sim$~20~x reduction from the room temperature value. The asymptotic divergence seen in these data at the higher frequencies is likely the result of an LC resonance feature of the test component~\cite{kalbitz}.The effect occurs at higher frequencies for lower temperatures.} 
\label{fig:Ch10_Cplot}
\end{figure}

The two nominally identical, multilayer ceramic 5XR 1812-size, 22~$\mu$F capacitors tested on Ch.~9 and Ch.~10 showed nearly identical performance, shown for example in Fig.~\ref{fig:Ch10_Cplot}.
\begin{table*}
\small
\caption{\label{tab:CapacitanceTable}Summary of capacitance values at different temperatures, as determined by averaging capacitance over a relatively constant portion of the low- to mid-frequency data, between $f_{min}$ and $f_{max}$. The reported uncertainty is two standard deviations as calculated from the data points included in the average and does not account for systematic errors propagated through the model. Parenthetical number pairs in the far left column indicate the open/short calibration channels used.}
\centering
\begin{tabular}{| r || r | l || r | l || r | l || r | l || l || r | r |}
\hline
Ch. & $C_{nom}$ & Tol. & $C_{300\,K}$ & $\delta C$ & $C_{12\,K}$ &$\delta C$ & $C_{360\,mK}$ & $\delta C$ & Units & $f_{min}$ (Hz) & $f_{max}$ (Hz)\\  
\hline
2(7/8) & 22 & $\pm$1\% & 21.3 & $\pm$0.9 & 20.3 & $\pm$0.8 & 20.7 & $\pm$0.6 & \textbf{pF} & 100 & 20k \\
\hline
5(11/12) & 10 & $\pm$1\% & 3.7 & $\pm$0.5 & 3.1 & $\pm$0.2 & 3.1 & $\pm$0.2 & \textbf{pF} & 100 & 20k \\
\hline
9(7/8) & 22 & $\pm$10\% & 21.9 & $\pm$1.0 & 1.38 & $\pm$0.09 & 0.951 & $\pm$0.018 & \textbf{$\boldsymbol{\mu}$F} & 0 & 200 \\
\hline
10(7/8) & 22 & $\pm$10\% & 21.6 & $\pm$0.9 & 1.37 & $\pm$0.09 & 0.940 & $\pm$0.018 & $\boldsymbol{\mu}$\textbf{F} & 0 & 200 \\
\hline
\end{tabular}
\end{table*}
At room temperature (300 K), the corrected impedance values for the 22 $\mu$F capacitors corresponded to a capacitance value well within the 10\% component tolerance. However, data collected at 12 K and 360 mK revealed a ``cold capacitance'' value roughly 20~x smaller than the nominal value. This significant drop in capacitance at low temperature helped explain the unexpected and elevated noise level recently observed with the {\SPLENDOR} two-stage amplifier, as reported in~\cite{anczarski}.

Previous studies on similar multilayer ceramic capacitors have found similar cryogenic behavior, especially for those capacitors which include high-$\kappa$ dielectric constant materials such as 5XR~\cite{homulle}. It has been established that high-$\kappa$ dielectric constant capacitors are generally of a ferroelectric nature and are optimized to maintain a large and stable dielectric constant only within a narrow range of rated operating temperatures which extends only down to $\sim$245 K~\cite{teyssandier,pan}. Although a decrease in capacitance at cryogenic temperatures has been documented, the magnitude of this effect varies by material and manufacturer~\cite{pan}. Therefore, it remains pertinent to directly measure the specific components to be used in a particular low-temperature detector system, so that the cryogenic behavior is known for that specific footprint size, component type, material, and manufacturer. 

\begin{figure}
\centering
\includegraphics[width=3.4 in]{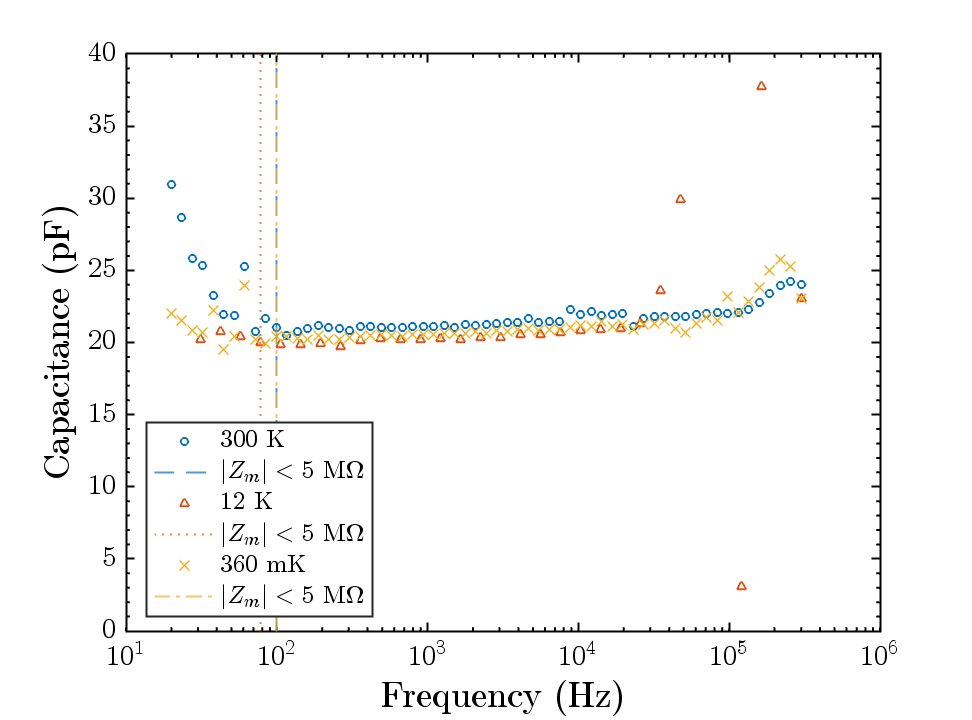}
\caption{``Open'' and ``short''-corrected capacitance measurements vs frequency for a nominally 22~pF thin film capacitor with an SiO$_2$/SiON dielectric at three test temperatures, 300~K (\textcolor{blue}{o}), 12~K (\textcolor{red}{$\triangle$}), and 360~mK (\textcolor{brown}{x}). The vertical dashed lines near 100 Hz indicate the frequency thresholds below which the measured impedances larger than 5 M$\Omega$. For calculating the average component impedance, only data for frequencies between 100 Hz and 20 kHz was included.}
\label{fig:Ch2_Cplot}
\end{figure}

Room temperature data for the 22~pF (nominal) test capacitor on Ch.~2 (see Fig.~\ref{fig:Ch2_Cplot}) yielded relatively constant capacitance values from 100~Hz to 20~kHz, with an averaged experimental value of 21.3 pF (95\% CL). The result was consistent with the 1\% tolerance specification provided by the manufacturer. At 12 K and 360 mK, the capacitance values were $\sim$ 2\% lower than the room temperature values, but still consistent with known tolerances.

\begin{figure}
\centering
\includegraphics[width=3.4 in]{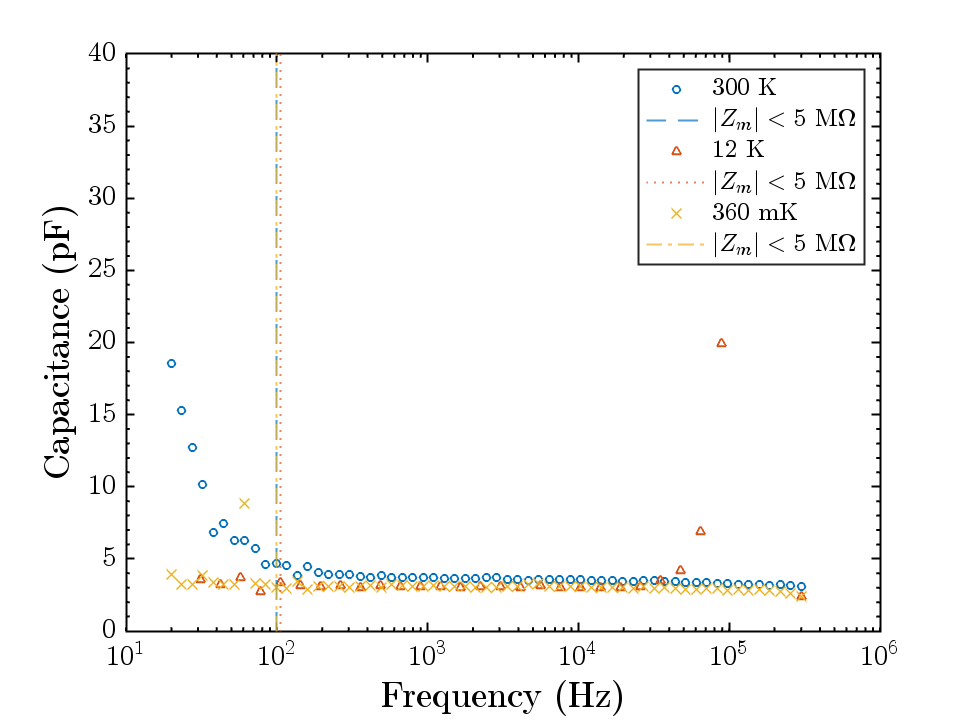}
\caption{``Open'' and ``short''-corrected capacitance measurements vs. frequency for a nominally 10~pF thin-film capacitor with an SiO$_2$/SiON dielectric, at three cryostat temperatures, 300~K (\textcolor{blue}{o}), 12~K (\textcolor{red}{$\triangle$}), and 360~mK (\textcolor{brown}{x}). The vertical lines indicate the frequency threshold ($\sim$100 Hz) below which the measured impedance was $>$5~M$\Omega$. For calculating the average component impedance, only data for frequencies between 100~Hz and 20~kHz were included.}
\label{fig:Ch5_Cplot}
\end{figure}

As shown in Fig.~\ref{fig:Ch5_Cplot}, the results for the nominally 10~pF capacitor on Ch.~5 were constant for frequencies between $\sim$400~Hz and 20~kHz. As occurred with the 22~pF capacitor, there was only a slight but possibly significant decrease in capacitance between 300~K and 12~K. However, unlike with the 22~pF capacitor, the 10~pF capacitor did not converge to its expected nominal value, but instead measured 3.1-3.7~pF. 
One possible explanation for this discrepancy is that our parasitics-correction model is simply not accurate for such low values of test component capacitance. Considering that the single-shot LCR meter reference measurements of the equivalent capacitance for the ``open'' circuit on Ch.~7 at 20~Hz gave values of $\sim$293~pF at 360 mK and $\sim$310 pF at 300~K, measuring a 10~pF component would mean precisely extracting a quantity $\sim$3\% of the magnitude of the measured parasitic background. Thus, especially in light of the universal LCR meter calibration offset noted in Section~\ref{Capacitors}, the 10~pF capacitor results should be taken as an indication that further scrutiny of the finer details of this circuit correction model is necessary to better address the possibility of small systematic errors propagating through the correction process, resulting in an offset uncertainty not accounted for in the purely statistical uncertainties listed in Table~\ref{tab:CapacitanceTable}.

We attribute the elevated effective capacitance at low frequency shown in Figures~\ref{fig:Ch2_Cplot}-\ref{fig:Ch5_Cplot}  to the likely presence of dielectric internal leakage resistance $R_p$ in parallel with the capacitor~\cite{teyssandier}. Modeling a real capacitor in this manner leads to the following expression for $C_x$:
\begin{equation}
    {C_x = \frac{1}{\omega\left(\frac{R_p^2C_{ideal}\omega}{1+R_p^2C_{ideal}^2}\right)}= \frac{1+R_p^2C_{ideal}^2\omega^2}{R_p^2C_{ideal}\omega^2}},
\end{equation}
where we have ignored any self-inductance of the capacitor, since the thin-film capacitors studied were designed for RF applications, and any self-resonant frequencies $\mathcal{O}$(GHz)~\cite{KyoceraAVX}, would be far above the frequency range under test. Writing the equation for $C_x$ in the more suggestive form:
\begin{equation}
    { C_x = C_{ideal}\left( 1+ \frac{1}{(2\pi fR_pC_{ideal})^2}\right)},
    \label{eq:CapFreqDep}
\end{equation}
makes clear how the capacitance value of a non-ideal capacitor can be significantly elevated at low frequencies, but converge to the expected, ideal value at higher frequencies, due to the $1/{f^2}$ dependence. While the temperature-related variation in capacitance in the intermediate frequency regime (where $C_x\approx C_{ideal}$ was negligible for the nominally 10~pF and 22~pF thin-film capacitors, both components exhibited greater low-frequency divergence at 300~K than at 12~K and 360~mK. Based on Eq.~\ref{eq:CapFreqDep}, this behavior suggests that $R_p$ is smaller at 300~K than at 12~K or 360~mK, consistent with the general trend that insulator resistance varies inversely with temperature~\cite{teverovsky}.

\subsection{Test Resistor Measurements}
\label{Resistors}

Four 100~M$\Omega$ thick-film test resistors from three different manufacturers were evaluated in this study. As shown in Table~\ref{tab:componentTbl}, the resistors on Ch.~2 (near the board edge) and Ch.~6 (an interior channel) were nominally identical, and were used in part to better understand board layout parasitics. 
For each test resistor, a component-isolated resistance was determined directly from the ``open'' and ``short''-corrected impedances through the relationship $R_x = \mathfrak{Re}(\mathbb{Z}_x)$, with $\mathbb{Z}_x$ defined in Eq.~\ref{eq:modelEqn1}.

\begin{figure}
\centering
\includegraphics[width=3.4 in]{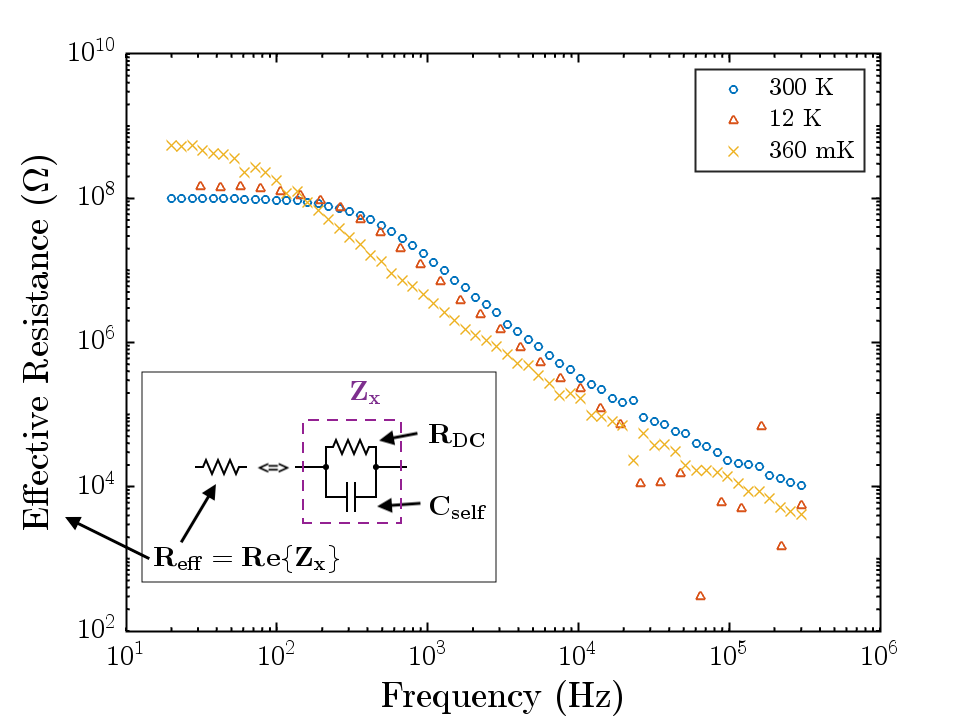}
\caption{Corrected effective resistance vs. frequency for the 100~M$\Omega$ (nominal) thick-film resistor on Ch.~6, at 300~K (\textcolor{blue}{o}), 12~K (\textcolor{red}{$\triangle$}) and 360~mK (\textcolor{brown}{x}). The decrease in $R_{eff}$ at higher frequencies is indicative of a non-zero self-capacitance for the resistor under test, in parallel with the ideal DC resistance. (See figure in-set.) Sampled frequencies used for this channel matched those on the ``open'' and ``short'' calibration channels, so frequency interpolation was not required for this analysis (see Table~\ref{tab:interpInfo}).}
\label{fig:Ch6_Rplot}
\end{figure}

As shown in Table~\ref{tab:ResistanceTable}, the nominal room temperature resistance of all four resistors was within the uncertainty range determined from the average near-DC measured resistance and standard deviation for data points below a 100 Hz upper-limit. The resistors on Ch.~4 and Ch.~6 exhibited less variance over this frequency range and had average values within 2\% of the nominal value. We observed a general tendency for the resistance to begin to drop-off at higher frequencies (see Fig.~\ref{fig:Ch6_Rplot}), indicative of parasitic self-capacitance in parallel with the test resistor. Such behavior has been well documented as the effective resistance of a resistor and capacitor in parallel~\cite{teyssandier}.

\begin{table*}
\small
\caption{\label{tab:ResistanceTable}Measurement summary for four, off-the-shelf, 100~M$\Omega$ resistors at different temperatures. The stated results were obtained by averaging resistance values over a low-frequency region $f<f_{max}$ where the resistance value remained nearly constant ($e.g$, see Fig.~\ref{fig:Ch6_Rplot}). The uncertainty values $\delta$$R$ shown correspond to $\pm$~2~$\sigma$ resistance values for the data points included in the average; the uncertainties do not explicitly include any systematic errors propagated through the model. Number pairs in parentheses in the far left column specify the ``open'' and ``short'' calibration channels used for each analysis.}
\centering
\begin{tabular}{| r || r | l || r | l || r | l || r | l || l || c |}
\hline
Ch. & $R_{nom}$ & Tol. & $R_{300\,K}$ & $\delta R$ & $R_{12\,K}$ &$\delta R$ & $R_{360\,mK}$ & $\delta R$ & Units & $f_{max}$ (Hz)\\  
\hline
1(7/8) & 100 & $\pm$1\% & 93 & $\pm$8 & 136 & $\pm$9 & 370 & $\pm$120 & \textbf{M$\boldsymbol{\Omega}$} & (100,100,50) \\
\hline
3(11/12) & 100 & $\pm$1\% & 97 & $\pm$6 & 134 & $\pm$16 & 460 & $\pm$340 & \textbf{M$\boldsymbol{\Omega}$} & (100,100,50) \\
\hline
4(11/12) & 100 & $\pm$5\% & 101 & $\pm$5 & 156 & $\pm$6 & 990 & $\pm$320 & \textbf{M$\boldsymbol{\Omega}$} & (100,100,50) \\
\hline
6(11/12) & 100 & $\pm$1\% & 98 & $\pm$4 & 142 & $\pm$7 & 480 & $\pm$120 & \textbf{M$\boldsymbol{\Omega}$} & (100,100,50) \\
\hline
\end{tabular}
\end{table*}

When cooled to 12 K, all four resistors increased in resistance by amounts within the range of $\sim$35-55\% of their 300 K values. This behavior is consistent with the behavior found by Novikov \textit{et al.}~\cite{novikov} for thick film Yageo RC0402 10-50 $\Omega$ resistors measured with an LCR meter and by Homulle \textit{et al.}~\cite{homulle} for a 33~k$\Omega$ thick film resistor when measured at cryogenic temperatures.
At 360 mK, the low-frequency effective resistance of all four off-the-shelf components was more than $\sim$3 times the nominal value, with the Ch.~4 resistor in particular exhibiting an increase in resistance close to an order of magnitude. Although the temperature coefficients reported by the manufacturers were only specified down to -55$^o$C, it is notable that the Ch.~4 resistor was rated at $\pm300$~ppm/$^o$C, whereas the other three resistors had coefficients of $\pm100$~ppm/$^o$C or less. 

\begin{figure}
\centering
\includegraphics[width=3.4 in]{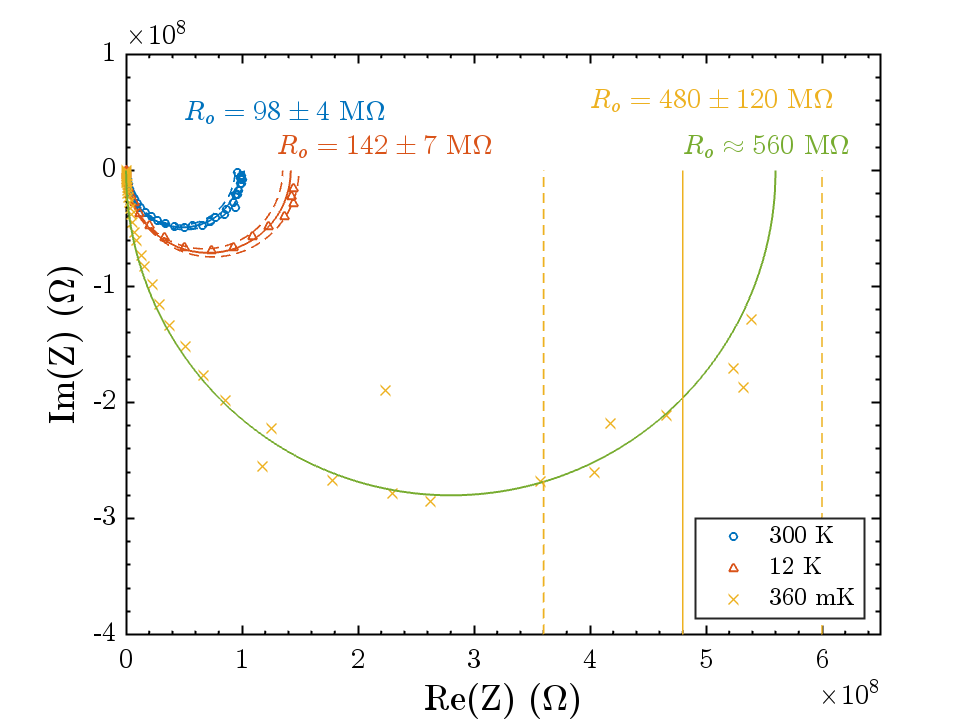}
\caption{Corrected complex impedance for the nominally 100 M$\Omega$ thick-film resistor on Ch. 6, measured at three temperatures. 
The solid-line semicircles shown have centers defined at $(x,y) = (\frac{R_o}{2},0)$, where $R_o$ is the predicted DC component resistance at each temperature. The values for $R_o$ for the 300~K data (\textcolor{blue}{o}) and 12~K data (\textcolor{red}{$\triangle$}) were taken from Table~\ref{tab:ResistanceTable}, along with their $\pm$~$2\sigma$ uncertainties. The (green) semicircle shown for the less-precise 360~mK data (\textcolor{brown}{x}) uses a fit-value of $R_o$ = 560~M$\Omega$ rather than the tabulated DC resistance value of 480 M$\Omega$. For comparison, the tabulated value of $R_o$~$\pm$~{$\delta$}R for the 360~mK data are represented by the vertical solid and dashed orange lines, respectively.}
\label{fig:Ch6_Nyquist}
\end{figure}

Without knowing any manufacturing details about these generic chip-resistors other than their quoted 300~K value, precision and size, it is beyond the scope of this paper to precisely model the internal physics of these resistors to explain their observed cryogenic behavior at 360~mK.
However, the relatively large uncertainty observed with the 360~mK resistance values, even when limiting the analysis to frequencies below 50 Hz, can be readily understood by examining the trajectory of $\mathbb{Z}_x$ in the complex plane. As seen for the Ch.~6 resistor in Fig.~\ref{fig:Ch6_Nyquist}, the 360~mK data points do not trace out a complete semicircle in the plane; the low-frequency (right-most) data do not extend along the semicircular path all the way up to the real axis. 

We note that a simple $\sim$10~Hz AC 4-wire resistance bridge could have been used to measure the effective DC resistances reported here. However, that approach would not have provided the opportunity to perform comprehensive, parasitic self-capacitance studies. Our method, by contrast, includes a simple, \textit{in situ} LCR meter calibration and capacitance-based analytical model that can readily provide accurate estimates of the self-capacitance of an unknown resistor.

\begin{figure}
\centering
\includegraphics[width=3.4 in]{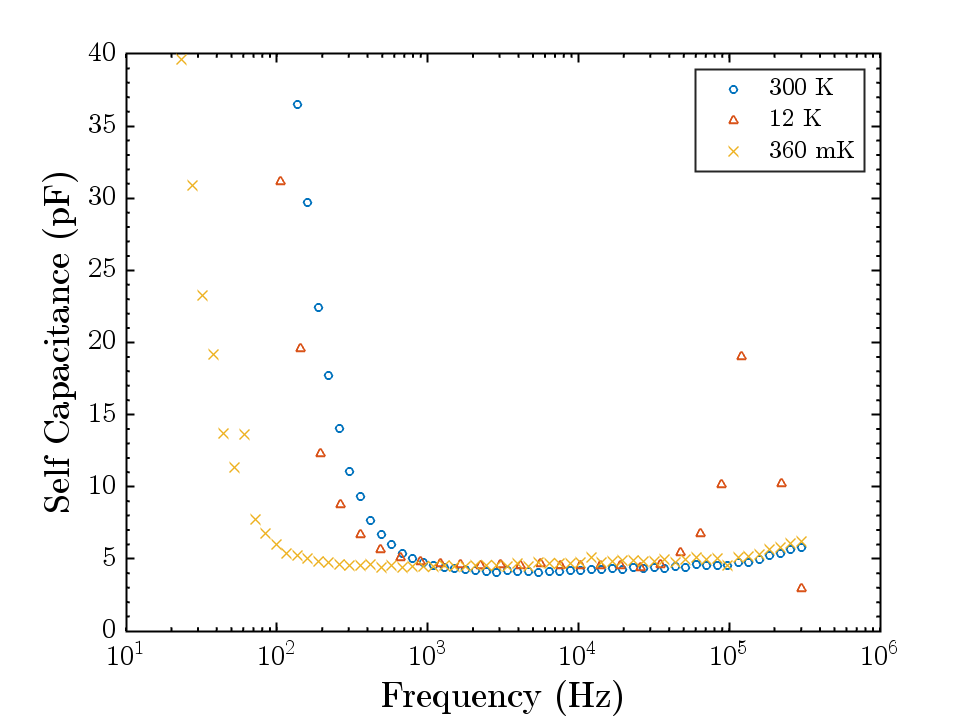}
\caption{Self-capacitance for the nominally 100~M$\Omega$ thick-film resistor on Ch. 6, as extracted from the imaginary component of the isolated component impedance, $\mathbb{Z}_x$. Data were taken at 300~K (\textcolor{blue}{o}), 12~K (\textcolor{red}{$\triangle$}) and 360~mK (\textcolor{brown}{x}). The estimated self-capacitance for this component is $\sim$5~pF, which stabilizes for the 360~mK data above 100~Hz (and above 1~khz for the higher temperature data). The self-capacitance value reported is subject to the same uncertainty considerations as was assumed for the 10~pF capacitor on Ch. 5.}
\label{fig:Ch6_Capacitance}
\end{figure}

For example, as shown in Fig.~\ref{fig:Ch6_Capacitance}, the 100~M$\Omega$ thick-film resistor on Ch.~6 was found to have a self-capacitance of $\sim$5~pF, which changes minimally with temperature. Based on the results reported for the 10~pF thin-film capacitor on Ch.~5, however, the model may have limited resolution at the pF scale, which is roughly 0.3\% of the magnitude of the known environmental parasitic capacitance. In order to evaluate the model accuracy within a testing regime better suited for the LCR meter, future studies could also examine pairs of components. For example, multiple (identical) 100~M$\Omega$ resistors mounted in parallel would have a lower combined effective resistance, but a higher combined self-capacitance. Such a set-up could be used to elevate the unknown self-capacitance above the existing $\sim$pF resolution threshold and ideally beyond the 22~pF benchmark set by the measurement of the capacitor tested on Ch.~2. 

\begin{figure}
\centering
\includegraphics[width=3.4 in]{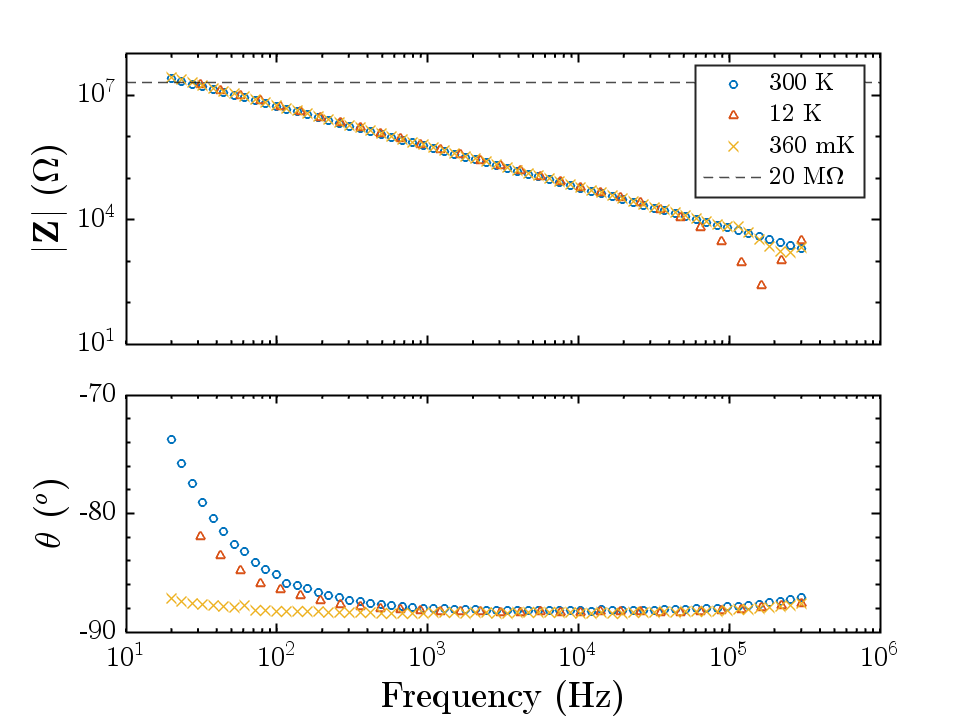}
\caption{As-measured impedance magnitude $\mathbb{Z}_m$ and phase data for the 100 M$\Omega$ resistor on Ch. 4 characterized at 300~K (\textcolor{blue}{o}), 12~K (\textcolor{red}{$\triangle$}) and 360~mK (\textcolor{brown}{x}). Parasitic contributions have not yet been removed. Notice that $|\mathbb{Z}_m|$ (top panel) does not exceed 30~M$\Omega$ at any frequency or temperature, and drops below the 20~M$\Omega$ upper-limit rating for the LCR meter range (dashed line) for frequencies above 27~Hz.}
\label{fig:Ch4_Zm}
\end{figure}

\begin{figure}
\centering
\includegraphics[width=3.4 in]{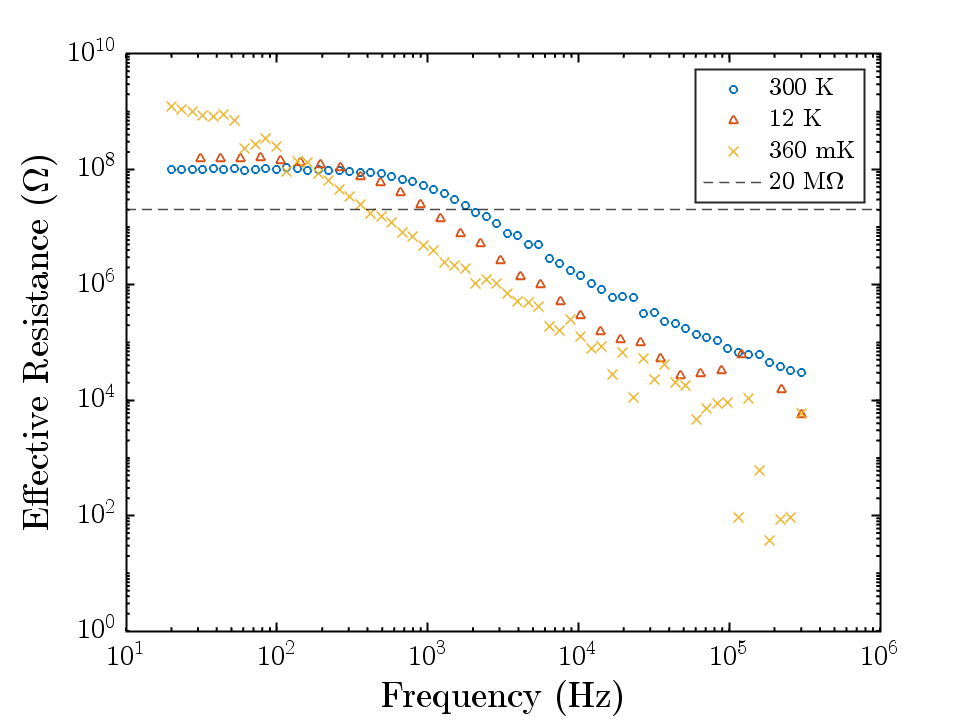}
\caption{Corrected effective resistance vs frequency at 300~K (\textcolor{blue}{o}), 12~K (\textcolor{red}{$\triangle$}) and 360~mK (\textcolor{brown}{x}) for the nominally 100 M$\Omega$ thick-film resistor on Ch.~4. Whereas the performance of the LCR meter is not rated above 20~M$\Omega$ (dashed line), the ``open'' and ``short'' calibration method and analytical model developed in this work can be used to accurately transform the raw data shown in Fig.~\ref{fig:Ch4_Zm} into actual component resistance values far beyond the quoted range capability of the LCR meter.}
\label{fig:Ch4_Rplot}
\end{figure}

It should be noted that the basic LCR meter used for this work was not rated for test component impedance values greater than 10-20~M$\Omega$ (see Fig.~\ref{fig:LCRspecs}). Nevertheless, we showed that our analytical model and $in situ$ calibration scheme can be used to take advantage of impedance-reducing parasitic effects, thus enabling accurate measurements of component resistances $>$100 M$\Omega$. This powerful ability to extend the useful range of the LCR meter is evident when comparing the Ch.~4, instrument-limited raw data values $\mathbb{Z}_m$ plotted in Fig.~\ref{fig:Ch4_Zm} to the same data plotted after processing as $\mathfrak{Re}(\mathbb{Z}_x)$, shown in Fig.~\ref{fig:Ch4_Rplot}. 

By contributing to an equivalent impedance with a reduced magnitude compared to the actual component impedance, the \textit{in situ} parasitic backgrounds effectively down-convert the component impedance into a range measurable by the LCR meter with greater accuracy than could be achieved by direct measurement of the component on its own. This effect is especially notable for the 360 mK data (as shown above), for which the \textit{post hoc} data transformation converts measured impedance magnitudes below 30 M$\Omega$ to component resistance values approaching 1 G$\Omega$, albeit with large uncertainty bounds.


\section{Conclusions}
\label{Conclusions}
We have developed a robust analytical model that can reliably extract sample resistance and capacitance values from complex data sets containing substantial and variable parasitic contributions. We also showed that the model can be used with a relatively low-cost LCR meter to perform scientifically meaningful, {\it in situ} cryogenic capacitor and resistor characterizations. 

Dedicated, on-board ``open" and ``short" correction channels readily enable measurements of isolated passive circuit components with \textit{post hoc} background subtraction of complex parasitics. Preliminary results confirmed a drop in capacitance by a factor of over 20 for two nominally 22~$\mu$F AVX multilayer ceramic capacitor with 5XR dielectric between operating temperatures of 300~K and 360~mK. Based on these results, we expect to significantly improve the performance of our existing two-stage {\HEMT} amplifier by more optimally selecting filter capacitors based on their (now) known ``cold capacitance'' values, rather than relying on nominal values provided by manufacturers.

Results for a 22~pF AVX thin-film capacitor with a SiO$_2$/SiON low-$\kappa$ dielectric suggest this type/brand of capacitor does not change appreciably with temperature, with the exception of an elevated insulator resistance at cryogenic temperatures. Further measurements for similar components should be conducted in the future to further probe the reproducibility of this transition. Additionally, results for a 10~pF AVX capacitor from the same series were stable across the full temperature range studied, but did not match the nominal value. Whether or not this is a limitation inherent to the model when measuring component values less than parasitic contributions remains to be seen and will be studied in future work.

Regarding the 100~M$\Omega$ thick-film resistors, our results generally converged to the nominal resistance value at 300~K. (See Fig.~\ref{fig:Ch4_Rplot}.) However, the ``DC'' resistance values were observed to increase with decreasing temperature. Measurements at 360~mK exhibited significantly elevated uncertainties, but had average magnitudes more than three times their nominal 300~K resistances. The Ohmite resistor in particular increased in resistance by roughly an order of magnitude between 300~K and 360~mK. Further measurements of these components, potentially in different configurations such as: (1)two 100~M$\Omega$ resistors in parallel on the same channel, or (2) a 100~M$\Omega$ resistor in parallel with an $\mathcal{O}$(nF) capacitor, for which an identical capacitor is also tested on its own on a separate channel, would be useful. Additionally, developing a procedure to fit corrected impedance data to curves in the complex plane could be explored as an alternative method for determining DC resistance and self-capacitance values with higher precision.

We did not explicitly seek out ``cryogenically-rated'' components for this development work. But, now that we have a robust method for accurately characterizing components over a very wide range of values,  we intend to do so for a large set of passive components that we expect will be useful for precision circuits operated at cryogenic temperatures. We also will extend the temperature range of such experiments down to $\sim$ 10~mK by mounting the existing test boards in a dilution refrigerator. Ultimately, the results could be compiled into a component database for reference by other members of the low-temperature detector community. By cataloging components by type and manufacturer, such a database would enable others to quickly locate components suitable for their specific designs based on $in situ$ behavior, rather than relying on nominal values that are known to be inaccurate in a cryogenic environment.

\section{Acknowledgments}
\label{Acknowledgments}
R.C. was responsible for the circuit modeling and data analysis that resulted in this paper. The raw data were obtained by R.C., with support from J.A. and others, using a $^3$He cryostat in the DMQIS labs at SLAC. B.Y. proposed the study, provided input on the analysis and interpretation of data, and with R.C. wrote the paper. All authors contributed to aspects of the cryogenic laboratory effort required to obtain the data used here. We thank Arran Phipps (CSUEB) for useful discussions about the two-stage SPLENDOR amplifier design, and Gary Sloan (SCU) for machining sample mounting parts. This research was supported in part by Geoff and Josie Fox Fellowships (R.C. and I.R.), a De Novo Fellowship (R.C.), the Barry Goldwater Scholarship and Excellence in Education Foundation (R.C.), and funding from a Lee and Seymour Graff Endowed Chair position at SCU (B.Y.). J.A. is supported by the Kavli Institute for Particle Astrophysics and Cosmology Chabollah Fellowship. SLAC is operated by Stanford University.


\end{document}

%% file: authors.tex
\author{R.J. Carpenter} 
\email[]{rcarpenter@scu.edu}
\affiliation{Department of Physics and Engineering Physics, Santa Clara University, Santa Clara, CA 95053, USA}

\author{J. Anczarski}
\affiliation{Department of Physics, Stanford University, Stanford, CA 94305, USA}
\affiliation{Kavli Institute for Particle Astrophysics and Cosmology, Stanford University, Stanford, CA 94035, USA}

\author{I. Rydstrom}
\affiliation{Department of Physics and Engineering Physics, Santa Clara University, Santa Clara, CA 95053, USA}

\author{A. Simchony}
\affiliation{Department of Physics, Stanford University, Stanford, CA 94305, USA}
\affiliation{SLAC National Accelerator Laboratory, Menlo Park, California 94025, USA}

\author{Z.J. Smith}
\affiliation{Department of Physics, Stanford University, Stanford, CA 94305, USA}
\affiliation{SLAC National Accelerator Laboratory, Menlo Park, California 94025, USA}

\author{N.A. Kurinsky}
\affiliation{Department of Physics, Stanford University, Stanford, CA 94305, USA}
\affiliation{SLAC National Accelerator Laboratory, Menlo Park, California 94025, USA}

\author{B.A. Young}
\email[]{byoung@scu.edu}
\affiliation{Department of Physics and Engineering Physics, Santa Clara University, Santa Clara, CA 95053, USA}

%% file: main.bbl
\begin{thebibliography}{99}  
\bibitem{baulieu} G. Baulieu, {$et.al,$} {\it J. Low Temp. Phys.} {\bf209}, 570-580 (2022).
\bibitem{juillard} A. Juillard, {$et.al,$} {\it J. Low Temp. Phys.} {\bf199}, 798-806 (2020).
\bibitem{phipps} A. Phipps, Y. Jin, and B. Sadoulet, {\it J. Low Temp. Phys.} {\bf176}, 470-475 (2014). 
\bibitem{anczarski} J. Anczarski, {$et.al,$} {\it J Low Temp Phys} {\bf214}, 256-262 (2024). 
\bibitem{lcrManual} \textit{B\&K Precision 300 kHz Bench LCR Meter User Manual}, B\&K Precision Corporation, Yorba Linda, CA, (2015).
\bibitem{kalbitz} R. Kalbitz, Wurth Elektronik, Appl. Note 109a [www.we-online.com], (2022).
\bibitem{lee} J.S. Lee, {$et.al,$} \textit{J Korean Ceram Soc} \textbf{49}, 475-483 (2012).
\bibitem{homulle} H. Homulle, S. Visser, B. Patra and E. Charbon,``Design techniques for a stable operation of cryogenic field-programmable gate arrays'', \textit{Rev. Sci. Instrum.} \textbf{89}, no.1, (2018).
\bibitem{teyssandier} F. Teyssandier and D. Prêle, \textit{Ninth International Workshop on Low Temperature Electronics}, (2010).
\bibitem{pan} M.-J. Pan, \textit{Cryogenics} \textbf{45}, 463-467 (2005).
\bibitem{novikov} I. Novikov, D. Volkhin and A. Vostretsov, \textit{IEEE 3rd International Conference on Problems of Informatics, Electronics and Radio Engineering} pp. 300-304, (2024).
\bibitem{KyoceraAVX} \textit{Accu-P Series Thin-Film RF/Microwave Capacitor Technology}, Kyocera AVX, [datasheet] TDS-RFM-0032.
\bibitem{teverovsky} A. Teverovsky, \textit{IEEE Trans. on Components, Packaging and Manufacturing Technology} \textbf{4}, 1169-1176, (2014).

\end{thebibliography}
